\documentclass[a4paper]{jpconf}
\usepackage{graphicx}
\usepackage{amsmath}
\usepackage{amssymb}
\usepackage{bm}
\newcommand{\rhoB}{\rho_{\scriptscriptstyle B}}
\newcommand{\muB}{\mu_{\scriptscriptstyle B}}
\newcommand{\sNN}{s_{\scriptscriptstyle NN}}
\newcommand{\MeV}{\mathrm{MeV}}
\newcommand{\GeV}{\mathrm{GeV}}
\newcommand{\TeV}{\mathrm{TeV}}
\newcommand{\Feff}{\mathcal{F}_\mathrm{eff}}
\begin{document}
\rightline{Preprint: YITP-15-127}
\vskip -4pt
\title{Approaches to QCD phase diagram;
effective models, strong-coupling lattice QCD, and compact stars}

\author{Akira Ohnishi}

\address{Yukawa Institute for Theoretical Physics, Kyoto University,
Kyoto 606-8502, Japan}

\ead{ohnishi@yukawa.kyoto-u.ac.jp}

\begin{abstract}
The outline of the two lectures given in "Dense Matter School 2015"
is presented.
After an overview on the relevance of the phase diagram 
to heavy-ion collisions and compact star phenomena,
I show some basic formulae to discuss the QCD phase diagram
in the mean field treatment of the Nambu-Jona-Lasinio model.
Next, I introduce the strong-coupling lattice QCD,
which is one of the promising methods to access the QCD phase diagram
including the first order phase boundary.
In the last part, I discuss the QCD phase diagram in asymmetric matter,
which should be formed in compact star phenomena.
\end{abstract}

\section{Introduction}
\begin{figure}[h]
\begin{center}
\begin{minipage}[c]{14pc}
\includegraphics[width=14pc]{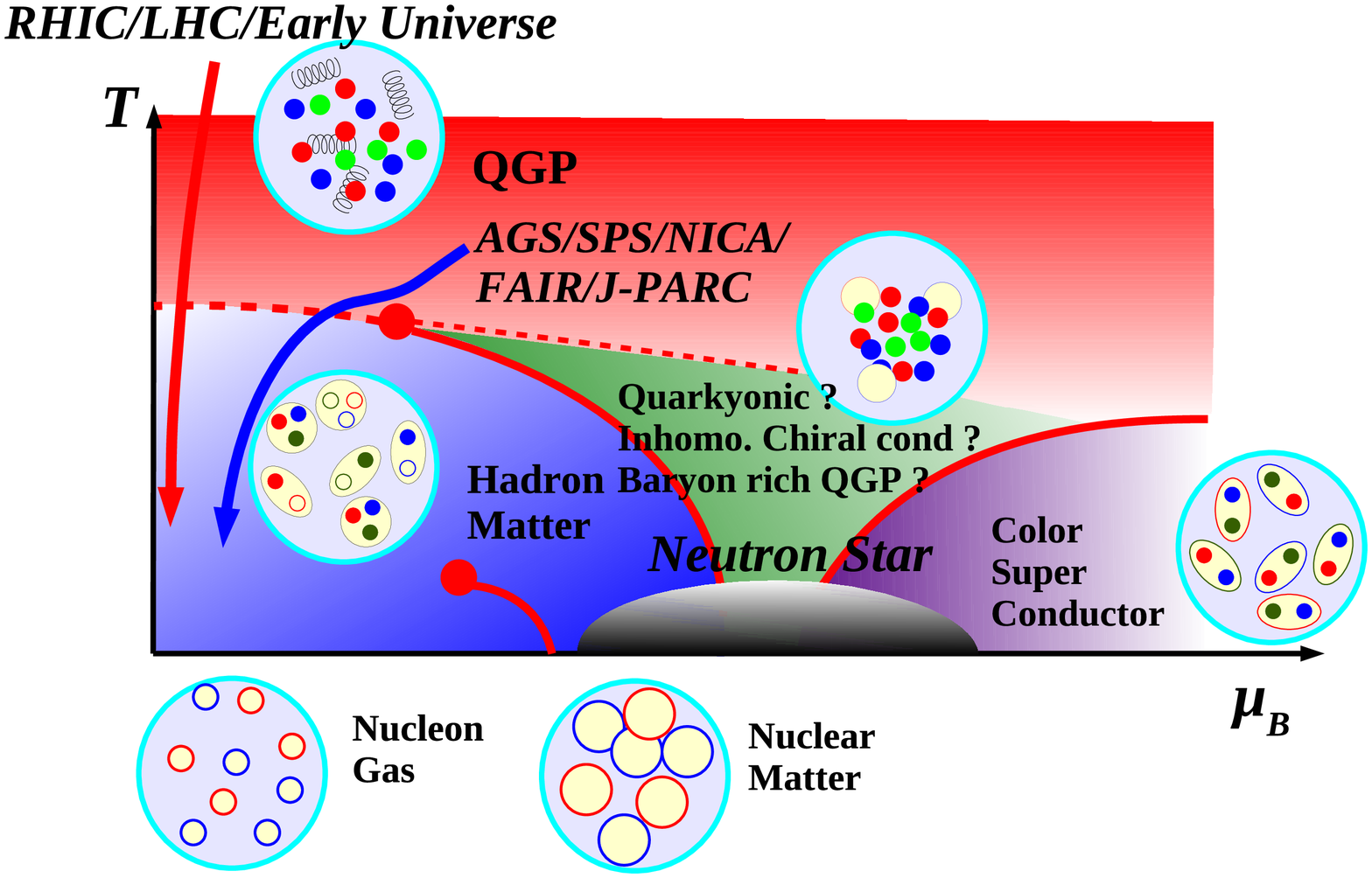}
\end{minipage}
\begin{minipage}[c]{14pc}
\includegraphics[width=14pc]{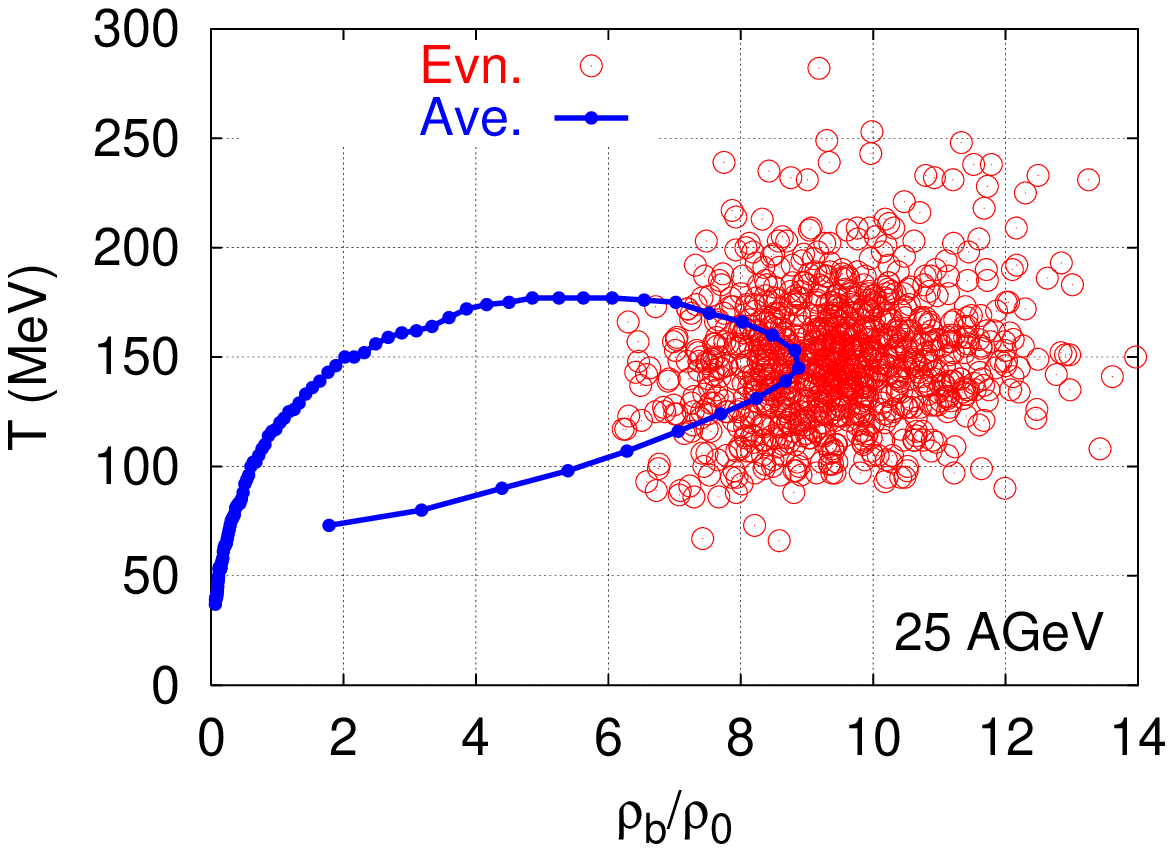}
\end{minipage}
\end{center}
\caption{\label{Fig:Sec1}
Left: Schematic QCD phase diagram~\cite{ISMD}.
Right: Temperature and density in 25 AGeV Au+Au collisions~\cite{JHF}.}
\end{figure}

QCD phase diagram is closely related to the history of our universe
and can be probed by using laboratory experiments
including heavy-ion collisions.
Recent developments in heavy-ion collision and compact star physics
have been providing hints to understand the QCD phase diagram
at finite temperature ($T$) and/or finite baryon density ($\rhoB$)
as schematically shown in the left panel of Fig.~\ref{Fig:Sec1}.
Heavy-ion collision data obtained at RHIC top energy ($\sqrt{\sNN}=200~\GeV$)
and LHC energy (run 1, $\sqrt{\sNN}=2.76~\TeV$) imply
the formation of strongly interacting matter consisting of quarks and gluons
at high $T$ and almost zero $\rhoB$~\cite{Stat}.
The phase transition at zero $\rhoB$ is known to be a crossover
from lattice QCD Monte-Carlo (LQCD-MC) simulations~\cite{crossover}.
Crossover nature of the transition justifies
the standard scenario of the homogeneous big bang nucleosynthesis.
In heavy-ion collisions at lower energies,
the rapidity gap between projectile and target nuclei becomes smaller.
Nucleons in the projectile and target lose their energies
and can get into the mid-rapidity region.
Thus the nuclear stopping gives rise to the formation of hot matter
at finite $\rhoB$.
In the right panel of Fig.~\ref{Fig:Sec1},
we plot the density and temperature reachable in Au+Au collisions
at $25 A\GeV$ ($\sqrt{\sNN}=7.1~\GeV$)
calculated in a hadron transport model, JAM~\cite{JAM}.
In a small central box of $1~\mathrm{fm}^3$,
the event averaged density may reach $9\rho_0$.
Unfortunately, the LQCD-MC simulations
suffer from the notorious sign problem,
which prevents us from obtaining precise phase boundary at finite $\rhoB$
from the first principles non-perturbative calculation.
Effective model calculations suggest the existence of the first order
phase transition boundary at finite density~\cite{CP}.
It is also possible that we have the region 
with inhomogeneous chiral condensate 
instead of a sharp phase boundary~\cite{inhomo}.
If the phase transition density is not very high,
we may have quark matter in the neutron star core,
whose density could reach $(5-10)\rho_0$.
Between the first order boundary (if exists) and the crossover transition,
there should be a critical point (CP)~\cite{CP},
where the transition is the second order.
Around CP, we expect large fluctuations of the order parameter
and other observables which couple with the order parameter~\cite{CPfluc}.
The location of CP determines the shape of the phase diagram,
and CP hunting is 
in progress in the current beam energy scan (BES) program at RHIC~\cite{BES}
and will be performed in the forthcoming facilities such as FAIR, NICA,
and heavy-ion beam facilities at J-PARC.

In this proceedings,
I give the outline of the two lectures given in "Dense Matter School 2015".
In the first lecture (before SQM),
I explained basic idea to describe the chiral phase transition
based on a mean-field treatment of a chiral effective model.
In the second lecture (during SQM),
I discussed how we can describe the QCD phase diagram
in the strong-coupling lattice QCD.
I also discussed the effects of isospin chemical potential
on the phase boundary,
and their implication to the compact star phenomena.

\section{QCD phase diagram in chiral effective models}

The QCD phase transition involves chiral and deconfinement transitions,
and
LQCD-MC results show that two transitions takes place at similar temperatures.
I here concentrate on the chiral transition.

Chiral symmetry is the symmetry of QCD with massless quarks.
The QCD Lagrangian is given as
\begin{align}
\mathcal{L}_\mathrm{QCD}
=&\bar{q}(i\gamma^\mu D_\mu - m)q - \frac12 \mathrm{tr} F^{\mu\nu}F_{\mu\nu}
\nonumber\\
=&\bar{q}_L(i\gamma^\mu D_\mu) q_L
+\bar{q}_R(i\gamma^\mu D_\mu)q_R
-(\bar{q}_L m q_R+\bar{q}_R m q_L)
- \frac12 \mathrm{tr} F^{\mu\nu}F_{\mu\nu}
\ ,
\end{align}
where $q_{L,R}=(1\mp\gamma_5)q/2$ are left- and right-handed quarks
and $F_{\mu\nu}$ is the field strength tensor of gluon.
When the quarks are massless ($m=0$),
$\mathcal{L}_\mathrm{QCD}$ is invariant under independent rotation
of $q_L$ and $q_R$ in the flavor space, $q_{L,R} \to q'_{L,R}=U_{L,R}q_{L,R}$,
where $U_{L,R}$ are $\mathrm{SU}(N_f)$ matrices.
In terms of hadrons, the chiral transformation mixes
$\sigma=\bar{q}q$ and $\pi^a=\bar{q}i\gamma_5\tau^aq$.
We should have a scalar meson having the same mass as pions
if the vacuum is also invariant under the chiral transformation.
In our world,
the chiral condensate has a finite value in vacuum,
$\langle{\bar{q}q}\rangle \not= 0$,
and the vacuum is not invariant under the chiral transformation.
Thus the symmetry of the QCD action with massless quarks
is spontaneously broken\footnote{$\mathrm{U}(1)_A$ is broken by the anomaly.},
$\mathrm{SU}(N_f)_L \otimes \mathrm{SU}(N_f)_R \otimes \mathrm{U}(1)_B \otimes \mathrm{SU}(N_c) \to \mathrm{SU}(N_f)_V \otimes \mathrm{U}(1)_B \otimes \mathrm{SU}(N_c)$.
This spontaneous breaking of chiral symmetry
in addition to the explicit breaking (finite bare quark mass)
gives rise to the mass difference of $\sigma$ and $\pi$ modes.
At high $T$ and/or $\rhoB$, spontaneously broken chiral symmetry is restored,
at least partially, {\em i.e.} the chiral transition takes place.

The chiral transition has been studied extensively
by using chiral effective models
such as the Nambu-Jona-Lasinio (NJL) model~\cite{NJL},
the Quark Meson (QM) model~\cite{QM},
and Polyakov loop extended versions of them such as PNJL~\cite{PNJL} 
and PQM~\cite{PQM} models.
These models have chiral symmetry,
and we expect that they could be obtained from QCD 
by integrating out high-momentum (short-range) degrees of freedom.
Let us consider the NJL model Lagrangian\footnote{We follow the notation
in Ref.~\cite{Yagi}},
\begin{align}
\mathcal{L}_\mathrm{NJL}
=&\bar{q}(i\gamma^\mu \partial_\mu - m_0 + \gamma_0\mu)q
+\frac{G^2}{2\Lambda^2}\left[
(\bar{q}q)^2+(\bar{q}i\gamma_5\bm{\tau}q)^2\right]
\ ,\\
\mathcal{L}_\mathrm{NJL}^{(E)}
=&\bar{q}(-i\gamma_\mu \partial_\mu + m_0 + i\gamma_4\mu)q
-\frac{G^2}{2\Lambda^2}\left[
(\bar{q}q)^2+(\bar{q}i\gamma_5\bm{\tau}q)^2\right]
\ .
\end{align}
In the second line,
we introduce the Euclidean action,
$\mathcal{L}^{(E)}=-\mathcal{L}(t=-i\tau,\gamma^0=-i\gamma_4,A^0=-iA_4)$,
where Euclidean coordinate and $\gamma$ matrices are
$(x_\mu)_E=(\tau=it,\mathbf{x})$
and
$(\gamma_\mu)_E=(i\gamma^0,\bm{\gamma})$, respectively.
The partition function in NJL is given as 
\begin{align}
\mathcal{Z}_\mathrm{NJL}
=&\int \mathcal{D}[q,\bar{q}] 
e^{
-\int_0^{1/T} d\tau \int d\mathbf{x} \mathcal{L}^{(E)}_\mathrm{NJL}
}
=\int \mathcal{D}[q,\bar{q},\sigma,\bm{\pi}] 
e^{
-\int d^4x \left\{
\bar{q}Dq
+ \frac{\Lambda^2}{2}(\sigma^2+\bm{\pi}^2)
\right\}
}
\nonumber\\
=&\int \mathcal{D}[\sigma,\bm{\pi}] 
\exp\left[-S_\mathrm{eff}(\sigma,\bm{\pi};T)\right]
\ ,\\
D=&-i\gamma\partial + m_0
+ i\gamma_4\mu + G(\sigma+i\gamma_5\bm{\tau}\cdot\bm{\pi})
\ ,\nonumber\\
S_\mathrm{eff}
=& - \log \det D + \int d^4x \frac{\Lambda^2}{2}\left[
\sigma^2(x)+\bm{\pi}^2(x)
\right]
\ .
\end{align}
In the second line, we have converted the four-Fermi interaction term in 
$\mathcal{L}^{(E)}_\mathrm{NJL}$
by using the Hubbard-Stratonovich transformation.

In order to calculate the partition function (or the free energy),
we need to obtain the determinant of the fermion matrix $D$
under the anti-periodic boundary condition in the temporal coordinate,
$q(1/T)=-q(0)$,
for a given configuration of the auxiliary fields, $(\sigma(x),\bm{\pi}(x))$,
and to integrate over the auxiliary fields.
We here work in the mean field approximation in order to illustrate
how the chiral transition occurs in a simple manner.
In the mean field approximation, we replace the auxiliary fields
with constant numbers, $(\sigma(x),\bm{\pi}(x)) \to (\sigma,0)$,
and $\sigma$ is chosen to minimize $S_\mathrm{eff}$.
Then the Fermion matrix becomes a Dirac operator with modified mass,
$D=-i\gamma\partial +i\gamma_4\mu+ m (m=m_0+G\sigma)$.
$D$ is diagonal in plane wave basis states,
$D=-\gamma_0(i\omega+\mu)+\bm{\gamma}\cdot\mathbf{k}+m$,
and the determinant is found to be\footnote{
The determinant of the usual (Minkowski) Dirac operator $D^{(M)}$
becomes zero for $E=\pm E_k$, then we get
$\det D^{(M)}=\prod(E^2-E_k^2)^{d_f/2}$.
By replacing $E\to i\omega+\mu$, we get $\det D$.}
$\mathrm{det} D = \prod_{n,\mathbf{k}}
((\omega_n-i\mu)^2+E_k^2)^{d_f/2}\ (E_k=\sqrt{\mathbf{k}^2+m^2})$,
where $d_f=4N_c N_f$ is the Fermion degrees of freedom
and $\omega_n=\pi T (2n-1)$ is the Matsubara frequency.
Now the free energy density is obtained as
\begin{align}
\Feff
=& - \frac{T}{V}\,\log \mathcal{Z} 
= \frac{\Lambda^2}{2}\sigma^2 
- \frac{T}{V} \sum_{n,\mathbf{k}}\,\log
\left((\omega_n-i\mu)^2+E_k^2\right)^{d/2}
\nonumber\\
=& \frac{\Lambda^2}{2}\sigma^2 
   -d_f \int \frac{d^3k}{(2\pi)^3}\left[
	\frac{E_k}{2} 
	+\frac{k^2}{3E_k}\,\frac12\left(
	 \frac{1}{e^{(E_k-\mu)/T}+1}
	+\frac{1}{e^{(E_k+\mu)/T}+1}
	\right)
	\right]
\nonumber\\
=& \frac{\Lambda^2}{2}\sigma^2
   - \frac{d_f}{2} \Lambda^4 I\left(\frac{m}{\Lambda}\right)
   - P^{(F)}
\ ,\label{Eq:FeffNJL}\\
I(x)=&\frac{1}{16\pi^2}\left[
\sqrt{1+x^2}(2+x^2)-x^4\log\frac{1+\sqrt{1+x^2}}{x}
\right]
\simeq
\frac{
1+x^2+\frac{x^4}{8}\left(1+4\log\frac{x}{2}\right)+\mathcal{O}(x^6)
}{8\pi^2}
\end{align}
From the first to the second line in Eq.~\eqref{Eq:FeffNJL},
Matsubara frequency sum is performed.
The first term in Eq.~\eqref{Eq:FeffNJL} comes from bosonization,
and the second and third terms come from the Fermion determinant
and show the zero point energy and thermal pressure, respectively.

\begin{figure}[h]
\begin{minipage}[c]{10pc}
\includegraphics[width=10pc]{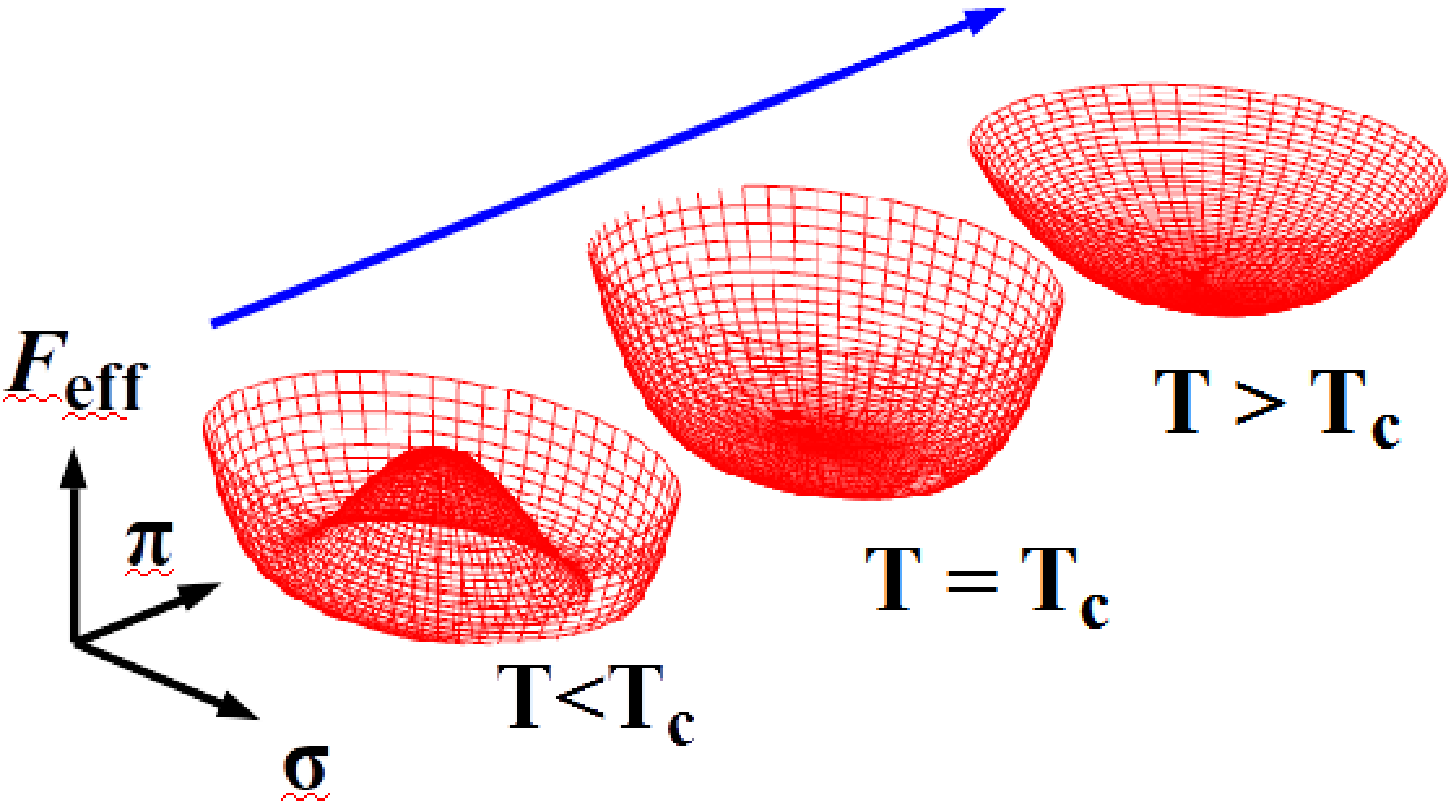}%
\end{minipage}
\begin{minipage}[c]{14pc}
\includegraphics[width=14pc]{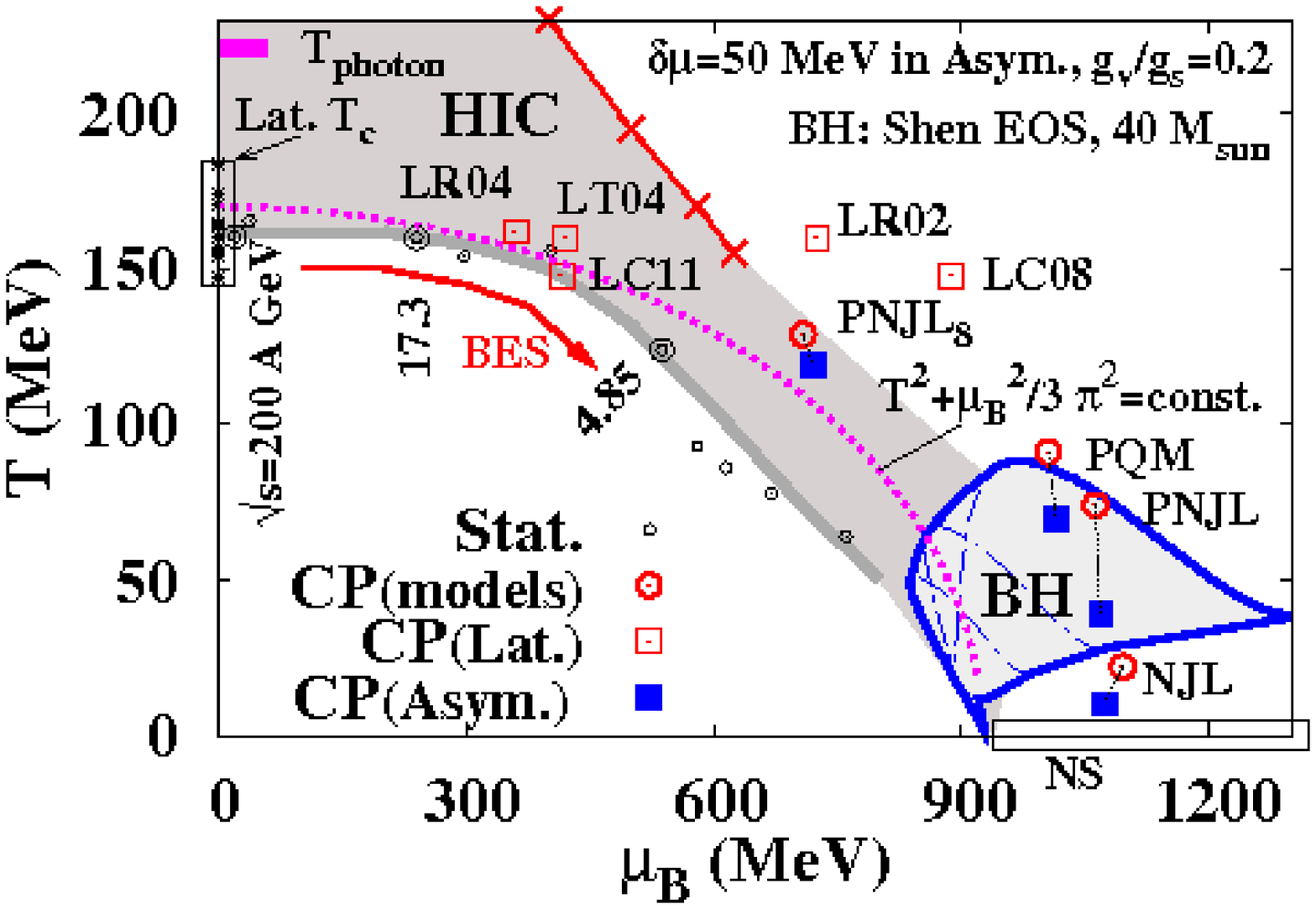}%
\end{minipage}
\begin{minipage}[c]{13pc}
~~~~~\includegraphics[width=13pc]{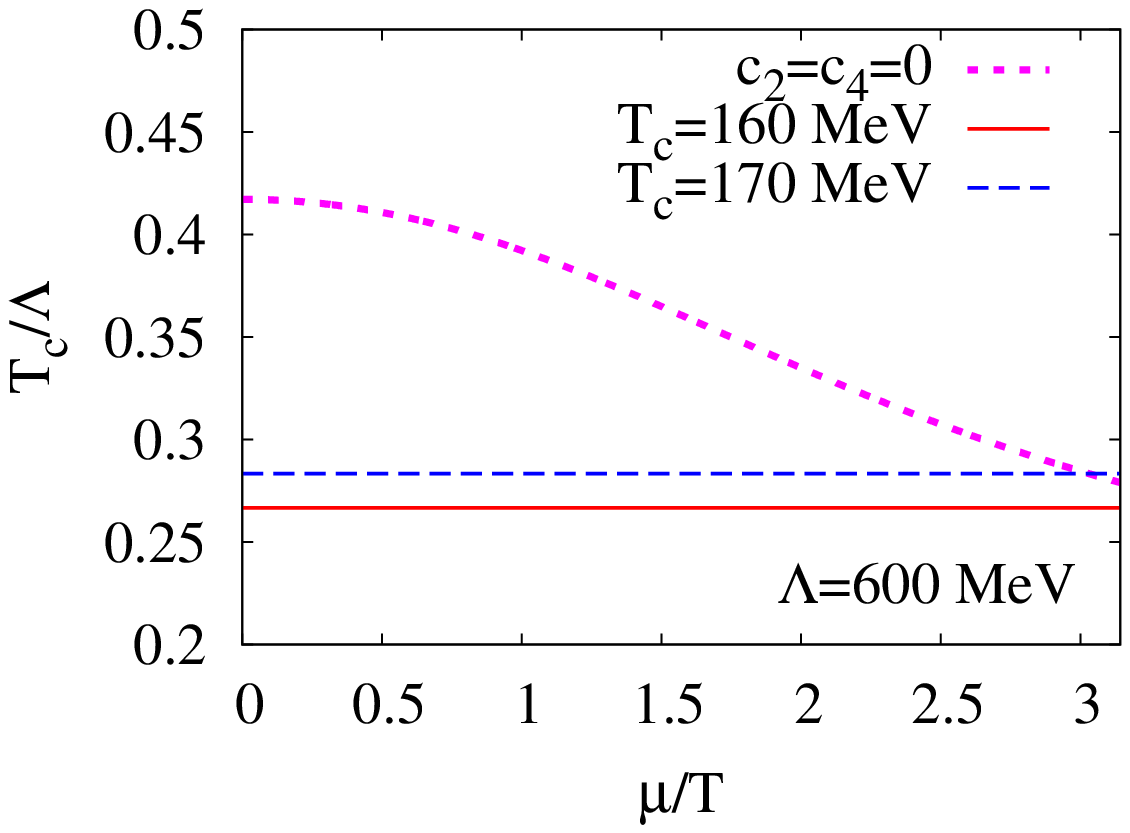}
\end{minipage}
\caption{\label{Fig:Sec2}
Left: Schematic view of the chiral phase transition.
Middle: Predicted critical points and observed chemical freeze-out 
points~\cite{ISMD}.
Right: Tricritical point condition.}
\end{figure}

The spontaneous symmetry breaking is understood
from the shape of $\Feff$ as a function of $\sigma$.
Let us consider the case in the chiral limit in vacuum
($m=0$, $T=0$ and $\mu=0$).
We can ignore the thermal contribution $P^{(F)}$,
and $\Feff$ can be written as
\begin{align}
\Feff/\Lambda^4 = - \frac{d_f}{16\pi^2} 
+ \frac{x^2}{2}\left[\frac{1}{G^2}-\frac{1}{G_c^2}\right]
+\mathcal{O}(x^4\log x)
\quad
(x=m/\Lambda ,\ G_c^2 = 8\pi^2/d_f)\ .
\end{align}
In the case $G>G_c$, the coefficient of $x^2$ is negative,
the $\Feff$ minimum appears at $x \not= 0$,
and the chiral condensate is non-zero in vacuum
as shown schematically in the left panel of Fig.~\ref{Fig:Sec2}.
Thus if the interaction is strong enough, 
$\sigma(=-\langle\bar{q}q\rangle)$ condensates
and constituent quark mass $m=G\sigma$ is generated.

Now let us discuss the chiral transition at finite $T$ and $\mu$,
where we need the mass dependence of the thermal pressure.
\begin{align}
P^{(F)}/d_f =& \frac78 \frac{\pi^2}{90}T^4 
+ \frac{\mu^2}{24}T^2 + \frac{\mu^4}{48\pi^2}
-\frac{m^2}{16\pi^2}\left[
\frac{\pi^2}{3}T^2 + \mu^2
\right]
\nonumber\\
-&\frac{m^4}{32\pi^2}\left[
\log\left(\frac{m}{\pi T}\right)-\frac34 + \gamma_E - H^\nu\left(\frac{\mu}{T}\right)
\right]
+\mathcal{O}(m^6)
\ ,\label{Eq:PrsTmu}\\
H^\nu(\nu)
=&\left(\frac{\nu}{\pi}\right)^2\sum_{l=1}^\infty
\frac{2}{(2l-1)[(2l-1)^2+(\nu/\pi)^2]}
=\sum_{k=1}^\infty (-1)^{k-1}\,
\left(\frac{\nu}{\pi}\right)^{2k}
\,\left(2-\frac{1}{2^{2k}}\right)\zeta(2k+1)
\ ,
\end{align}
where $\gamma_E= 0.5772156649\cdots$ is the Euler's constant.
The first three terms in Eq.~\eqref{Eq:PrsTmu} are the massless quark
contributions (Stefan-Boltzmann law),
and other terms show mass dependence of the pressure~\cite{Kapusta}.
We again consider the chiral limit ($m_0=0$),
then we can expand $\Feff$ in the Taylor series of $m$ as
\begin{align}
\Feff(m;T,\mu)
=&\Feff(0;T,\mu)+\frac{c_2(T,\mu)}{2}\,m^2 
+ \frac{c_4(T,\mu)}{24}\,m^4 + \mathcal{O}(m^6)
\ ,\\
c_2(T,\mu)=&\frac{d_f}{24}\left[
T^2+\frac{3}{\pi^2}\mu^2-T_c^2
\right]
\quad\left(
T_c^2=\frac{3\Lambda^2}{\pi^2}\left[1-\frac{G_c^2}{G^2}\right]
\right)
\ ,\label{Eq:c2}\\
c_4(T,\mu)=&\frac{3d_f}{4\pi^2}\left[
\gamma_\mathrm{E}-1-\log\left(\frac{\pi T}{2\Lambda}\right)
-H^\nu\left(\frac{\mu}{T}\right)
\right]
\ .
\end{align}
We assume that the interaction is strong enough
and chiral symmetry is spontaneously broken in vacuum,
then $T_c^2>0$.
At $\mu=0$, $c_2$ changes its sign at $T=T_c$,
then the chiral restored state ($m=G\sigma=0$) is realized in equilibrium
at $T>T_c$.
This chiral transition should occur at $T<\sqrt{3}\Lambda/\pi$,
which is less than the cutoff.

The second order phase boundary is given by the condition $c_2(T,\mu)=0$
provided that $c_4>0$,
and is found to be $T^2+3\mu^2/\pi^2=T_c^2$.
If this boundary reaches the $\mu$ axis,
the critical baryon chemical potential is evaluated to be
$N_c\mu_c=\pi N_cT_c/\sqrt{3}$,
which amounts to be around the nucleon mass $N_c\mu_c\simeq 925~\MeV$
for $T_c \sim 170~\MeV$.
This elliptically shaped phase boundary roughly explains
the chemical freeze-out points probed in heavy-ion collisions,
as shown in the middle panel of Fig.~\ref{Fig:Sec2}.

The second order boundary may be terminated by the tricritical point (TCP),
where $c_2=0$ and $c_4=0$ is simultaneously satisfied.
For the free energy density in Eq.~\eqref{Eq:PrsTmu},
the TCP condition is expressed as
\begin{align}
\frac{T_c}{\Lambda}
=\frac{2\sqrt{1+3\nu^2/\pi^2}}{\pi}\,e^{\gamma_E-1-H^\nu(\nu)}
\quad (\nu=\mu/T)\ .
\label{Eq:TCP}
\end{align}
In the right panel of Fig.~\ref{Fig:Sec2},
dotted line shows the right hand side of Eq.~\eqref{Eq:TCP},
and solid and dashed lines show $T_c/\Lambda$ values with $\Lambda=600~\MeV$
for $T_c=160~\MeV$ and $170~\MeV$, respectively.
For $T_c=170~\MeV$, the TCP condition is satisfied
around the conversion radius of the high-temperature expansion, $|\mu/T|=\pi$. 

When the quark mass is finite,
the transition at small $\mu$ becomes the crossover,
and TCP becomes CP if the first order boundary exists.
The location of CP is sensitive to the details of the model.
The NJL model in the present simple treatment
predicts a large $\mu/T$ value of TCP, $\mu/T \gtrsim \pi$, 
as discussed above.
When the confinement and/or the 8-Fermi interaction effects
are taken into account~\cite{8Fermi}, CP temperature increases.
Some lattice QCD calculations predict $T_\mathrm{CP}$ similar to $T_c(\mu=0)$.
We need experimental data on CP and the first order phase transition,
and/or some breakthrough in lattice QCD at finite density
in order to pin down the location of CP
and to understand the structure of the QCD phase diagram.

\section{QCD phase diagram in strong-coupling lattice QCD}

The lattice QCD Monte-Carlo (LQCD-MC) simulations is
the non-perturbative and first-principles method of QCD,
and have been applied to various observables
such as hadron masses, hadron-hadron interactions,
and QCD thermodynamics at $\mu=0$.
At finite density, however, the fermion determinant becomes complex
and it becomes difficult to perform precise calculation at large $\mu$.
There are many methods proposed so far in order to avoid the sign problem.
The strong-coupling lattice QCD is one of the promising methods.

In this section, I briefly introduce the basic idea of the lattice QCD
and the sign problem.
Next, I explain the strong-coupling lattice QCD.

\subsection{Lattice QCD}

We consider here a lattice QCD action 
for color $\mathrm{SU}(N_c)$
with one species of unrooted staggered fermion
in the $d(=3)+1$ dimensional Euclidean spacetime
with $N_\tau$ temporal and $L$ spatial lattice sizes.
\begin{align}
S_\mathrm{LQCD}
=&\bar{\chi}D\chi + S_G
\label{Eq:LQCD-ActionB}\\
=&\frac12 \sum_x \left[
\bar{\chi}_x U_0(x)\,e^\mu \chi_{x+\hat{0}}
-\bar{\chi}_{x+\hat{0}} U_0^\dagger(x)\,e^{-\mu} \chi_{x}
\right]
\nonumber\\
+&\frac12 \sum_{x,j} \eta_j(x) \left[
\bar{\chi}_x U_j(x)\,e^\mu \chi_{x+\hat{j}}
-\bar{\chi}_{x+\hat{j}} U_j^\dagger(x)\,e^{-\mu} \chi_{x}
\right]
+m_0 \sum_x \bar{\chi}_x \chi_x
\nonumber\\
+&\frac{2N_c}{g^2}\sum_\mathrm{plaq.} \left[
1 - \frac{1}{N_c} \mathrm{Re}\,\mathrm{tr}\,U_{\mu\nu}(x)
\right]
\ .
\label{Eq:LQCD-Action}
\end{align}
Spacetime is discretized on the lattice,
$(\tau,x,y,z)=(a i_\tau, a i_x, a i_y, a i_z)$, 
where $a$ is the lattice spacing and $i_\mu$ is an integer,
$1 \leq i_\tau \leq N_\tau$ and $1 \leq i_j \leq L\ (j=1,2,3)$.
In this proceedings, we adopt the lattice unit, $a=1$.
%
%
Quarks are represented as anti-commuting Grassmann numbers
on the lattice sites, $\chi_x^a$ 
(color index $a$ is suppressed in Eq.~\eqref{Eq:LQCD-Action}).
Since the square of a Grassmann number is zero,
we just need to define $\int d\chi=0$ and $\int \chi d\chi=1$
to evaluate the path integral;
the former is {\em an anti-commuting constant} which should be zero,
and the latter is a commuting constant which can be defined as unity.
Using these definitions,
we find $\int \exp(\bar{\chi}A\chi) d\bar{\chi} d\chi=\det A$.
The link variable $U_\mu(x)=P\exp\left[ig\int_x^{x+\hat{\mu}}dx A_\mu(x)\right]$
is an $N_c\times N_c$ matrix function of the gluon field
and defined on a link $(x,x+\hat{\mu})$.
The gauge transformation of the quark and link variables are given as
$\chi_x \to V_x\chi_x$,
$\bar\chi_x \to \bar\chi_xV_x$,
and 
$U_\mu(x)\to V_xU_\mu(x)V_{x+\hat{\mu}}$.
Then the combination $\bar\chi_xU_\mu(x)\chi_{x+\hat{\mu}}$ is gauge invariant.
The staggered factor $\eta_j(x)=(-1)^{x_0+\cdots+x_{j-1}}$ represents
$\gamma_j$ for staggered fermions,
then the first two lines in Eq.~\eqref{Eq:LQCD-Action} leads to
the quark action in the continuum QCD.
The trace of the plaquette
$U_{\mu\nu}(x)=U_\mu(x)U_\nu(x+\hat{\mu})U_\mu^\dagger(x+\hat{\mu}+\hat{\nu})U_\nu^\dagger(x+\hat{\nu})$
is also gauge invariant.
The loop integral $\oint dx A$ appears in the exponent 
in $U_{\mu\nu}$ and generates the {\em rotation} $F_{\mu\nu}$,
as deduced from the Stokes theorem in the $\mathrm{U}(1)$ gauge case.

The partition function of lattice QCD is given as
\begin{align}
\mathcal{Z}_\mathrm{LQCD}
=& \int \mathcal{D}[\chi,\bar{\chi},U_0,U_j]\,e^{-\bar{\chi}D\chi - S_G}
= \int \mathcal{D}[U_0,U_j]\,\det D\,e^{-S_G}
\ .
\end{align}
The integrand $\det U\,\exp(-S_G)$ is regarded as a statistical weight
for a configuration of link variables.
The Fermion matrix $D$ has $\gamma_5$ Hermiticity,
and its determinant is real at $\mu=0$,
\begin{align}
\left[\gamma_5 D(\mu) \gamma_5\right]^\dagger=D(-\mu^*)
\ ,\quad
\left[\det D(\mu)\right]^*=\det\left[D(-\mu^*)\right]
\ .
\end{align}
Then it is possible to perform the Monte-Carlo integral
by regarding $\det D\,\exp(-S_G)$ as a statistical weight
for a configuration of $U$ {\em at zero $\mu$}.
At finite $\mu$, $\det D$ becomes a complex number,
and a na\"ive probability interpretation fails down.
There are many attempts to avoid the sign problem.
Many of these methods are useful for $\mu/T < 1$,
while it is difficult to
perform the Monte-Carlo simulation in the larger $\mu/T$ region,
where we may expect the first order phase transition.

\subsection{Strong-coupling lattice QCD}

In the strong coupling region, it is possible to explore the phase diagram
including the first phase transition boundary.
%
The strong-coupling lattice QCD is a method
based on the $1/g^2$ expansion of the partition function,
and has a long history of study.
In 1974, Wilson showed in the first work on the lattice gauge field theory
that the Wilson loop follows the area law
in the strong coupling limit~\cite{Wilson},
\begin{align}
\langle{W(C=L\times N_\tau)}\rangle
= \frac{1}{\mathcal{Z}} \int \mathcal{D}[U] W(C)\,e^{-S_G}
\to N_c \left(\frac{1}{g^2N_c}\right)^{LN_\tau} \quad (g^2 \to \infty)
\ ,
\end{align}
where $LN_\tau$ is the area of the Wilson loop
as shown in the left panel of Fig.~\ref{Fig:Sec3}.
We need at least $LN_\tau$ plaquettes in order to kill
all unpaired link variables.
By using the one-link integral formula shown later in Eq.~\eqref{Eq:OneLink},
each plaquette generates a factor $1/N_cg^2$ to the Wilson loop.
The Wilson loop $W(C)=\tr P \exp(i\oint_C A_\mu dx_\mu)$
is related with the potential between heavy quarks,
$\langle{W(C)}\rangle=\exp(-V(L)N_\tau)$.
The area law tell us that $V(L)$ contains a linear confining potential,
$V(L)=L\log(g^2N_c)$.
Thus color is found to be confined in the strong coupling region.
This confining feature was confirmed by Creutz
in the first lattice QCD Monte-Carlo simulation in 1980~\cite{Creutz}.
The observed string tension is found to connect
the strong coupling region and the weak coupling region smoothly.
Strong-coupling expansion with higher order terms
was performed by M\"unster~\cite{Munster},
and the MC results are well reproduced.

Strong-coupling lattice QCD with quarks was first explored
by Kawamoto and Smit~\cite{Kawamoto}.
It was shown that chiral symmetry is spontaneously broken
in the strong-coupling limit~\cite{Kawamoto,Aoki}.
This idea is extended to finite $T$ and 
$\mu$~\cite{SCL-Tmu,SCL-PD,SC-PD-plaq,MDP,MDP-PD,AFMC,MDP-PD-plaq,AFMC-plaq}.
The phase transition in the strong-coupling and chiral limit
is found to be the second order at $\mu=0$
and the first order at $T=0$,
in the mean field treatment~\cite{SCL-Tmu,SCL-PD}
and with fluctuation effects~\cite{MDP-PD,AFMC}.
Effects of plaquette terms at finite $1/g^2$ on the phase diagram
are also studied~\cite{SC-PD-plaq,MDP-PD-plaq,AFMC-plaq}.

\subsection{Phase diagram at strong coupling}

The plaquette term $S_G$ in the lattice QCD action Eq.~\eqref{Eq:LQCD-Action}
is proportional to $1/g^2$,
and can be treated as perturbation in the strong coupling region.
Especially, we can neglect $S_G$ in the strong coupling limit,
then it is possible to integrate out the link integral independently.
We can perform the one-link integral by using the formulae,
\begin{align}
\int dU U_{ab} =0
\ ,
\int dU U_{ab} U^\dagger_{cd}=\frac{1}{N_c}\,\delta_{ad}\,\delta_{bc}
\ ,
\int dU U_{ab} U_{cd} U_{ef} 
= \frac{1}{N_c!}\,\varepsilon_{ace}\,\varepsilon_{bdf}
\ ,
\ldots
\label{Eq:OneLink}
\end{align}
The second formula gives an effective action terms
after integrating out spatial link variables,
\begin{align}
S_\mathrm{eff}^\mathrm{(SCL)}
=& S_F^{(t)} - \frac{1}{4N_c}\sum_{x,j} M_x M_{x+\hat{j}} + m_0 \sum_x M_x
\quad (M_x=\bar{\chi}_x \chi_x)\ ,
\label{Eq:SeffSCL}
\end{align}
where $S_F^{(t)}$ is the temporal hopping term of quarks
shown in the first term in Eq.~\eqref{Eq:LQCD-Action}.
This effective action contains the nearest neighbor four-Fermi interaction.
The third one-link integral formula generates six-Fermi interaction
(baryon hopping term),
which is in the sub-leading order in the large dimensional ($1/d$) 
expansion~\cite{largeD}.
Let us consider the case with large spatial dimension, $d \gg 1$.
In order to keep the four-Fermi interaction term finite at $d \to \infty$,
quark field should scale as $\chi \propto d^{-1/4}$
to compensate a factor $d$ from the sum over $j$.
The six-quark terms are $\mathcal{O}(d^{-1/2})$ for $N_c=3$,
and we ignore them in the later discussion.

\begin{figure}[h]
\begin{center}
\includegraphics[width=10pc]{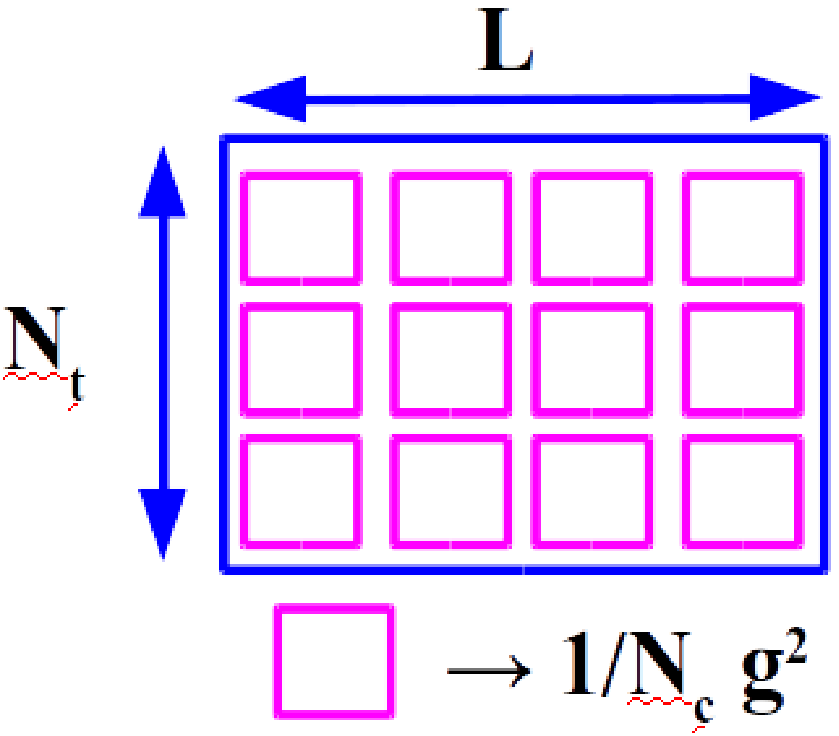}~~%
~\includegraphics[width=12pc]{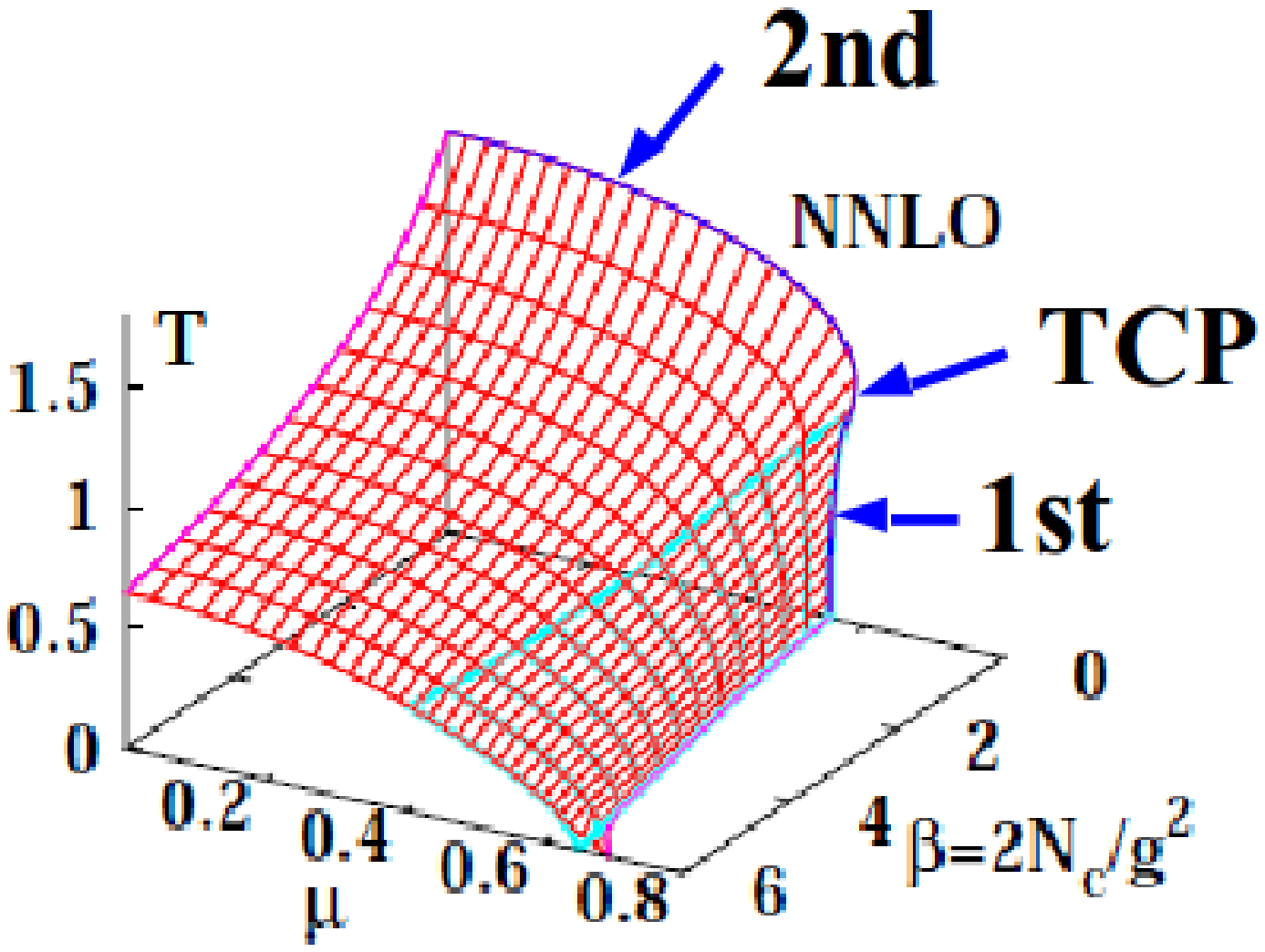}%
~\includegraphics[width=12pc]{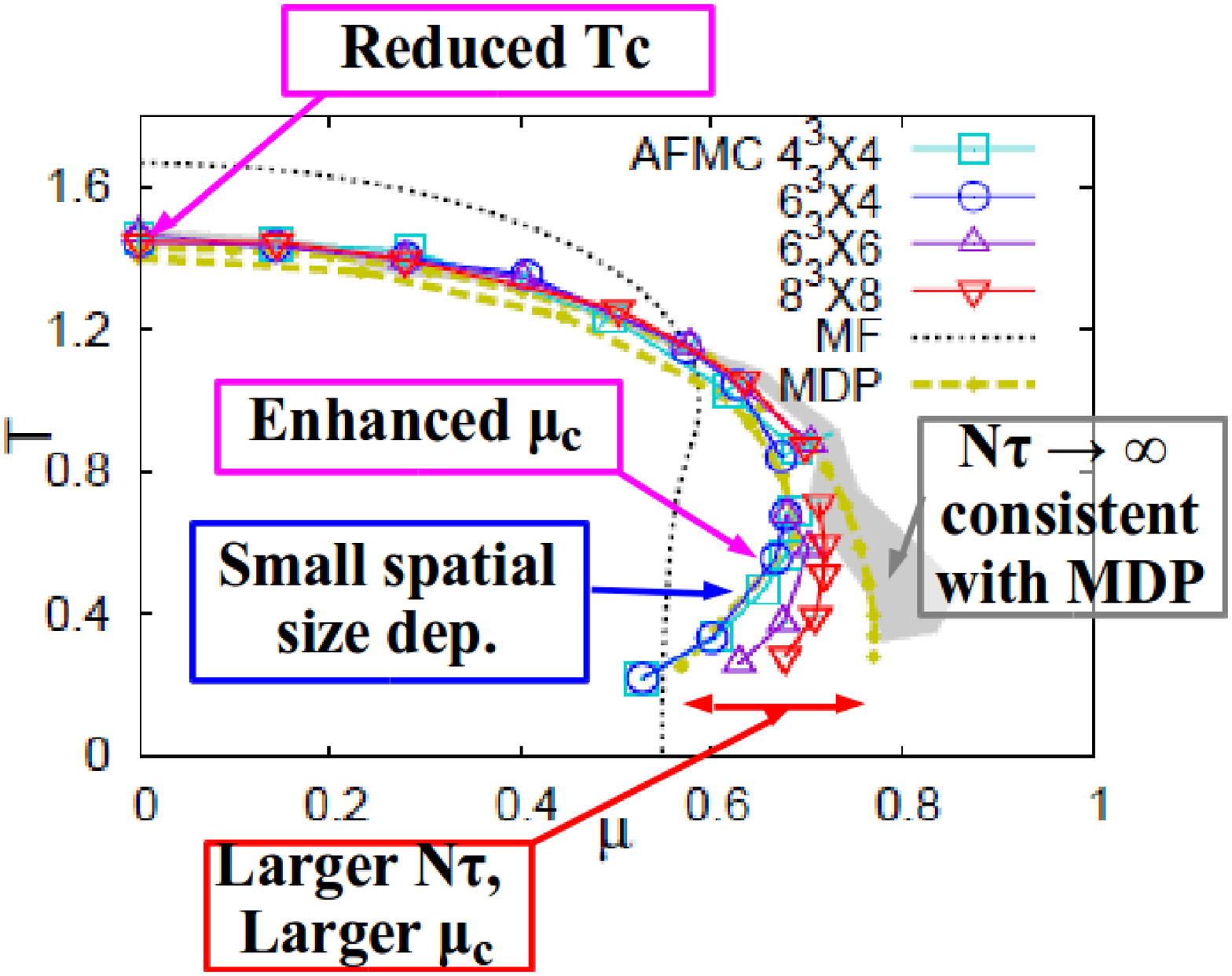}%
\end{center}
\caption{\label{Fig:Sec3}
Left: Schematic figure of the Wilson loop.
Middle: Phase diagram as a function of inverse coupling, $\beta_g=2N_c/g^2$~\cite{SC-PD-plaq}.
Right: Phase diagram in the strong coupling limit~\cite{AFMC}.
}
\end{figure}

Once an action with four-Fermi interaction is given,
there are several standard methods to solve the quantum many-body problem.
One of these methods is the bosonization of the action
followed by the mean field approximation and the Grassmann integral.
The phase diagram in the strong coupling limit has been studied
in the mean field treatment of the effective action 
given in Eq.~\eqref{Eq:SeffSCL}~\cite{SCL-PD,SC-PD-plaq}.
The obtained phase diagram has similar features of that predicted
in chiral effective models; 
the second order (crossover) transition at low $\mu$,
the first order transition at large $\mu$,
and the existence of TCP (CP)
in (off) the chiral limit,
as shown 
in the left panel of Fig.~\ref{Fig:Sec3} (the $\beta=2N_c/g^2=0$ surface).
This similarity of the phase diagrams owes to the symmetry of the theory.
$S_\mathrm{eff}^\mathrm{(SCL)}$ in Eq.~\eqref{Eq:SeffSCL}~\cite{SCL-PD,SC-PD-plaq}
in the chiral limit ($m_0=0$)
is invariant under the $\mathrm{O}(2)$ chiral transformation
($\chi_x \to e^{i\varepsilon_x\alpha/2}\chi_x$,
$\bar{\chi}_x \to e^{i\varepsilon_x\alpha/2}\bar{\chi}_x$),
and anomaly is not realized in the strong coupling region.
From the universality argument,
the system with O(2) symmetry without anomaly 
shows the second order phase transition.
This can be proxy for the system with O(4) symmetry with anomaly,
as in the $N_f=2+1$ QCD in the chiral limit for $u$ and $d$ quarks.

The above QCD phase diagram is not satisfactory in several points;
mean field approximation (no fluctuation effects),
the strong coupling limit ($1/g^2=0$),
only the leading order of the $1/d$ expansion,
and one species of unrooted staggered Fermion.
There are some recent progress on the fluctuation effects.
One of the methods to exactly evaluate the partition function
is the Monomer-Dimer-Polymer (MDP) simulation~\cite{MDP,MDP-PD}.
We integrate out all link variables in the strong coupling limit,
and we find that the partition function is represented
as a sum of weights for monomer-dimer-polymer configurations.
Some of the MDP configurations have negative weights,
but the weight cancellation is weak.
In MDP, we can also include the baryon hopping term in the sub-leading order
of the $1/d$ expansion.
Another method proposed so far is 
the auxiliary field Monte-Carlo (AFMC) method~\cite{AFMC},
which is a straightforward extension of the mean field treatment.
After bosonizing the effective action,
we keep the auxiliary fields as they are
without putting them as constant mean fields,
and integrate over the auxiliary fields using the Monte-Carlo method.
%
We show the phase diagram obtained in these two methods 
in the strong coupling and chiral limit
in the right panel of Fig.~\ref{Fig:Sec3}.
Anisotropic lattice ($a_\tau \not= a_s$) is adopted in these calculations.
Green dashed lines show the phase boundaries
in MDP with $N_\tau=4$ and $\infty$.
Solid lines with symbols show the phase boundaries in AFMC
for $4^4, 6^3\times4, 6^4$ and $8^4$ lattices.
Compared with the mean field results,
fluctuation effects reduces the transition temperature at $\mu=0$,
and enhance the hadron phase at large $\mu/T$.
For larger $N_\tau$, the transition $\mu$ becomes larger.
These features are common in the two methods,
and the phase boundaries are found to agree with each other.

Plaquette term effects (finite $1/g^2$ effects) are also investigated
in the mean field treatment~\cite{SC-PD-plaq}.
In the middle panel of Fig.~\ref{Fig:Sec3},
we show the phase diagram as a function of $1/g^2$.
The transition temperature decreases with increasing $1/g^2$
and is found to be consistent with the standard
MC calculation results in the strong coupling region
($\beta_g=2N_c/g^2 \lesssim 4$),
when we take account of the $1/g^2$ terms
and the Polyakov loop terms ($(1/g^2)^{N_\tau}$).
Calculations including both plaquette and fluctuation effects
have been performed recently.

An interesting application of the strong-coupling lattice QCD
is the net-baryon number cumulants across the phase boundary~\cite{IMO2015}.
Around the phase transition, we expect large fluctuations of order parameters.
The order parameter at CP is a linear combination of the chiral condensate
and the baryon density~\cite{FujiiOhtani},
then larger baryon number fluctuations are expected
in heavy-ion collisions at $\sqrt{\sNN}=(5-20)~\GeV$,
where the formed matter is considered to go around CP.
Net-baryon number cumulants
\begin{align}
\chi^{(n)}=\partial^n(P/T^4)/\partial (N_c\mu/T)^n=\langle(\delta B)^n\rangle_c
\end{align}
show the baryon number fluctuations,
and higher-order cumulants are known to be more sensitive
to the criticality~\cite{Stephanov}.
Recently observed data in the beam energy scan show that 
the cumulant ratio $\kappa\sigma^2=\chi^{(4)}/\chi^{(2)}$
shows a non-monotonic behavior
as a function of $\sqrt{\sNN}$, which might signal the existence of CP.
Since the susceptibility ($\chi^{(2)}$) diverges at CP,
it is dangerous to rely on the Taylor expansion in $\mu/T$ from $\mu=0$.

We have recently obtained the net-baryon number cumulants
in the strong coupling and chiral limit.
The cumulant ratio $\kappa\sigma^2$ shows oscillatory behavior
as a function of $T$ for a given $\mu/T$,
and the negative $\kappa\sigma^2$ region appears along the phase boundary
at $\mu/T\gtrsim 0.2$.
The lattice size dependence shows diverging behavior
in agreement with the scaling function analysis~\cite{FKRS2011}.
It is important to study the finite mass effects
in order to understand the observed non-monotonic behavior of $\kappa\sigma^2$.

\section{QCD phase diagram of isospin-asymmetric matter and compact stars}

Now let us turn to the compact star phenomena.
In the previous sections, we have discussed the phase diagram
of symmetric matter, where chemical potential is the same for all quarks.
Dense matter probed in compact star phenomena, however, is not symmetric.
Before the core collapse, the electron-to-baryon ratio\footnote{
$Y_e$ in this proceedings should be interpreted as
the ratio of electric charge of hadrons and baryon number ($Y_Q$)
when muons and hyperons are taken into account.}
in the iron core is around $Y_e \sim 0.46$,
and it decreases to around $Y_e \sim 0.3$ as the electron capture proceeds.
In neutron stars, it is further small, $Y_e \lesssim 0.1$.
In binary neutron star mergers,
$Y_e$ at high density is similar to that in neutron stars,
while $Y_e$ has variety in the low density region
leading to the r-process nucleosynthesis~\cite{Sekiguchi}.
If we ignore hyperons, charge neutrality requires the same number of protons
and electrons, then the proton fraction (proton-to-baryon ratio)
$Y_e=Y_p=Z/(Z+N)=(0-0.46)$ in compact star phenomena.
It is well known that symmetric ($N\sim Z$) and neutron-rich ($N>Z$) nuclei 
have different properties such as the binding energy, nuclear radius,
and excitation spectra.
Then we should ask ourselves. 
How does the QCD phase diagram evolve as a function of the proton fraction ?

\begin{figure}[h]
\begin{center}
\includegraphics[width=10pc]{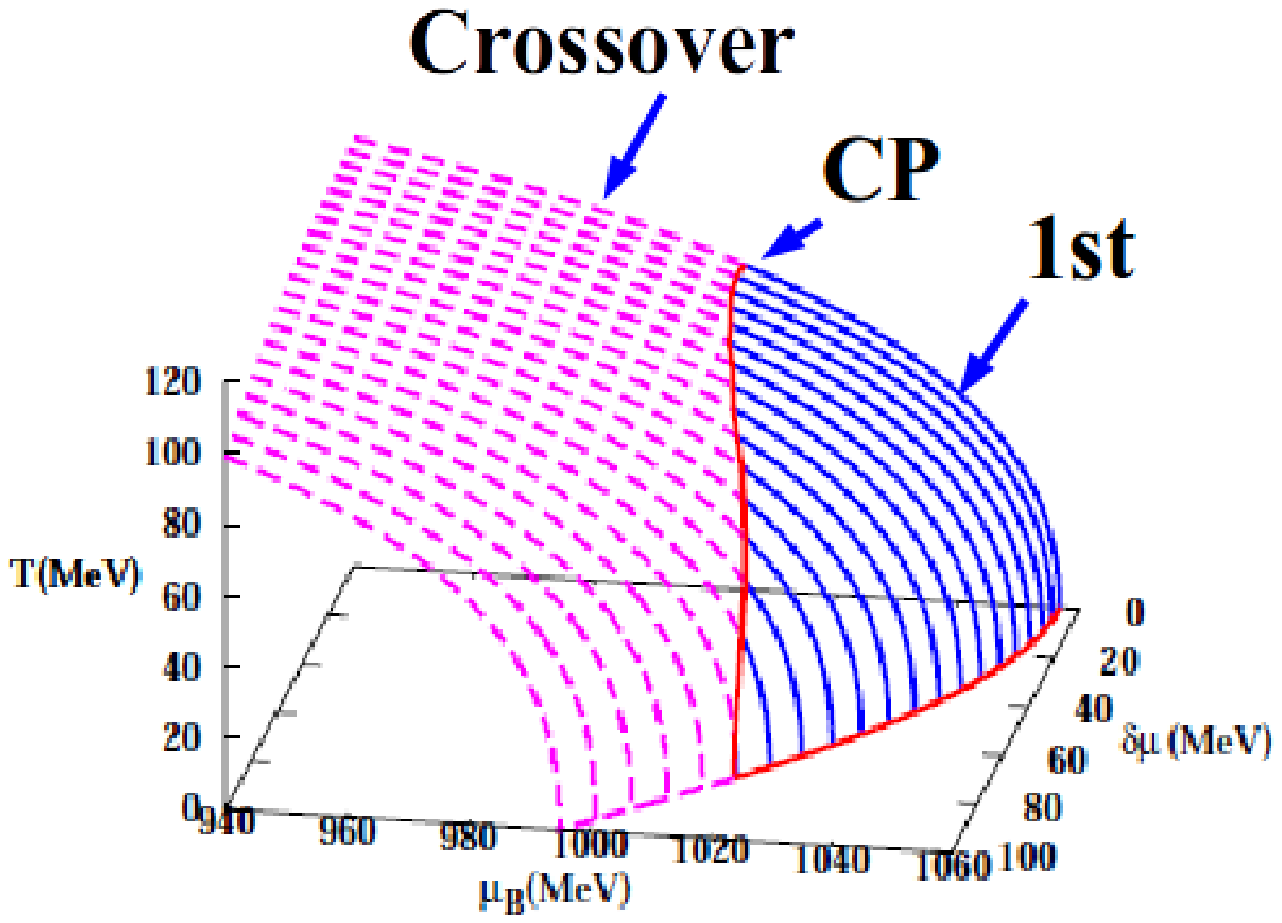}~~%
~\includegraphics[width=12pc]{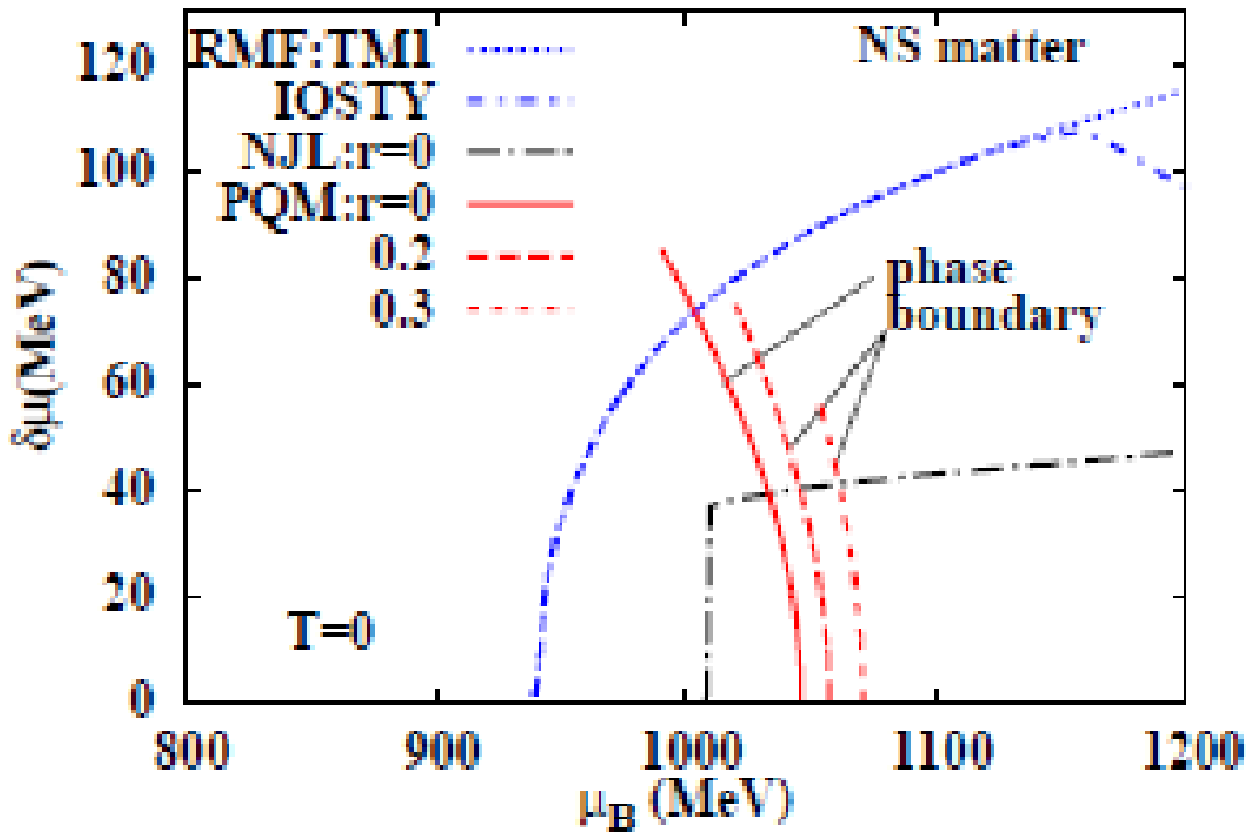}%
\end{center}
\caption{\label{Fig:Sec4}
Left: Phase diagram as a function of isospin chemical potential~\cite{Ueda}.
Right: First order phase boundary in comparison with isospin chemical potential
in neutron star matter~\cite{Ueda}.}
\end{figure}

Chiral effective model approach is again useful
to explore the QCD phase diagram in asymmetric matter.
We consider the Polyakov-loop extended Quark Meson (PQM) model
with vector interaction,
\begin{align}
\mathcal{L}_\mathrm{PQM}
=& \bar{q}\left[ i\gamma^\mu D_\mu
- g_\sigma (\sigma + i \gamma_5 \bm{\tau}\cdot\bm{\pi})\right] q
+ \frac12 \partial^\mu\sigma\partial_\mu\sigma
+ \frac12 \partial^\mu\bm{\pi}\partial_\mu\bm{\pi}
-U_\sigma(\sigma,\bm{\pi})-U_\Phi(\Phi,\bar{\Phi})
\nonumber\\
-&g_v \bar{q}\gamma^\mu(\omega_\mu+\bm{\tau}\cdot\bm{R}_\mu) q
-\frac14 \omega_{\mu\nu}\omega^{\mu\nu}
-\frac14 \bm{R}_{\mu\nu}\cdot\bm{R}^{\mu\nu}
+\frac12 m_\omega^2\omega_{\mu}\omega^{\mu}
+\frac12 m_\rho^2\bm{R}_{\mu}\cdot\bm{R}^{\mu}
\ .
\end{align}
In addition to quarks, and $\sigma$ and $\pi$ mesons,
we consider $\omega$ and $\rho$ (represented by $\bm{R}$) vector mesons.
$\Phi$ and $\bar{\Phi}$ are the Polyakov and anti-Polyakov loops.
The meson potential is chosen to be the sum of
the zero-point energy of quarks, polynomial of $\sigma^2+\bm{\pi}^2$,
and explicit breaking term $-h\sigma$.
The Polyakov-loop potential is tuned to fit to the lattice results.
The vector coupling of quarks are not well fixed,
then we regard it is a free parameter.

In the left panel of Fig.~\ref{Fig:Sec4}, 
we show the QCD phase diagram in $(T, \muB, \delta\mu)$ space,
where $\delta\mu=(\mu_d-\mu_u)/2=(\mu_n-\mu_p)/2$
is the isospin chemical potential.
We have applied the mean field approximation
with an assumption that the pions do not condensate,
according to the $s$-wave $\pi{N}$ repulsion argument~\cite{swave}.
Compared with the phase diagram of symmetric matter ($\delta\mu=0$),
the transition temperature decreases with increasing $\delta\mu$.
CP temperature also decreases with $\delta\mu$,
and eventually CP disappears at $\delta\mu=(70-80)~\MeV$.
The reduction of the transition temperature can be understood in part
by using the high-temperature expansion formula in Eq.~\eqref{Eq:PrsTmu}
and the curvature $c_2$ in Eq.~\eqref{Eq:c2} in the chiral limit.
Since $u$- and $d$-quark chemical potentials are given as
$\mu_u=\mu-\delta\mu$ and $\mu_d=\mu+\delta\mu$,
we find $c_2=d_f/24\left[T^2+3(\mu^2+(\delta\mu)^2)/\pi^2-T_c^2\right]$
for the $N_f=2$ NJL model in the chiral limit,
and the second order phase transition temperature decreases 
with increasing $\delta\mu$ for given $(T,\mu=\muB/3)$.
We expect that the effects of the vector coupling, the Polyakov loop,
and finite quark mass are less sensitive to $\delta\mu$.
This 3D phase diagram is similar to that
obtained in the functional renormalization group (FRG) method~\cite{FRG},
while it contradict to the effective model results 
with pion condensation~\cite{pion}.

Shrinkage of the hadron phase and the first order phase transition boundary
could be relevant to compact star phenomena.
First, the disappearance of the first order phase boundary
at large $\delta\mu$ suggest the possibility of the crossover transition
in the neutron star core.
In the middle panel of Fig.~\ref{Fig:Sec4},
we show isospin chemical potential of neutron star matter
in $\beta$ equilibrium 
as a function of baryon chemical potential
calculated by using a relativistic mean field (RMF) model, TM1,
in comparison with the first order phase boundary at $T=0$ in PQM
with the vector-scalar coupling ratio $r=g_v/g_\sigma=0, 0.2$ and $0.3$.
In the cold neutron star core,
$\delta\mu$ is calculated to reach around 100 MeV.
When the vector coupling is not very small ($r>0.2$),
the first order phase boundary disappears below $\delta\mu=80~\MeV$
and the chiral phase transition could be the crossover.
If this is the case, the softening of the equation of state
from the first order phase transition is weakened.
This supports the crossover scenario~\cite{Masuda}
to support the $2M_\odot$ neutron stars~\cite{MassiveNS}.

Another interesting possibility is the critical point sweep
during the dynamical black hole formation~\cite{CPsweep}.
We consider here the numerical results in \cite{Sumi2006};
Gravitational collapse of a 40 $M_\odot$ star is simulated
using the $\nu$ radiation 1D (spherical) hydrodynamics
with the Shen EOS~\cite{ShenEOS}.
In failed supernovae, core collapse of massive stars directly form black holes.
Temperature can reach $T\sim 90~\MeV$ at off-center by the shock heat,
which is above the CP temperature of asymmetric matter 
in many of effective models.
In the center, temperature is relatively low ($T<40~\MeV$)
while the density is high ($\muB \sim 1300~\MeV$ just before the black hole
formation).
Thus there is a possibility that matter at the center goes through
the first order phase transition boundary below CP,
the off-center goes above CP,
and some of the mass-shell hit CP.
We expect large baryon number fluctuations of that mass-shell,
but we have not yet found possible observational signals, unfortunately.

\section{Summary}

In this proceedings, I gave the outline of my two lectures
in the Dense Matter 2015.
We expect the existence of QCD phase transition
and the critical point from chiral effective model studies.
This point is discussed based on the Nambu-Jona-Lasinio model.
When interaction is strong enough, chiral symmetry is
spontaneously broken in vacuum.
Chiral symmetry should be restored at high temperature.
Density effect reduces the 4-th order coefficient in constituent quark mass
which is proportional to the chiral condensate in the chiral limit.
Thus we can expect the existence of the critical point
and the first order transition at high density.
In the lecture I also explained some technical part, 
such as the Matsubara sum, Hubbard-Stratonovich transformation,
and high-temperature expansion.
Since the first principle calculation of QCD has difficulties
at finite densities, we need studies using effective models,
approximate treatment of QCD, and of course, experiments.

In the second lecture, I explained other two approaches
to the QCD phase diagram.
The first one is the strong-coupling lattice QCD.
While we have the sign problem in lattice QCD at finite $\mu$,
the phase diagram study is on going using various ideas.
I have shown recent results based on the strong-coupling lattice QCD.
Smaller weight cancellation allow us to study phase transition at high density.
Phase diagram in the strong coupling limit has been confirmed,
through the agreement of MDP and AFMC results.
While the strong-coupling limit is the opposite limit of the continuum limit,
cumulant ratio calculation would be valuable to understand
the appearance of the criticality in finite volume.

The second one is the phase diagram of asymmetric matter and compact stars.
Compact stars are also good laboratories of dense matter.
In neutron stars, supernovae, black hole formation,
and binary neutron star mergers,
we expect the formation of dense and isospin asymmetric matter.
With the first order boundary (and CP) and isospin chemical potential,
there are many ways of realizing phase transition in compact star phenomena.

Finally, I would like to emphasize that
dense matter is "terra incognita", where there are many unsolved problems.
In heavy-ion collisions at $\sqrt{\sNN} =(5-10)\GeV$,
we expect formation of highest baryon density matter,
whose density may exceed $5\rho_0$.
In equilibrium, this would be above the transition density.
In compact star phenomena,
hydro simulations with hadronic matter EOS suggest
the formation of dense matter ($(4-5)\rho_0, \muB \sim 1300~\MeV$),
which is above the transition density in many effective models.
We need more experimental, observational,
and theoretical works to explore dense matter.

\ack
The author would like to thank David Blaschke and other members in JINR
for their hospitality.
The author also thanks Kenta Kiuchi and Yuichiro Sekiguchi
for preparing the figure for the lecture.
This work is supported in part by the Grants-in-Aid for Scientific Research 
from JSPS (Nos.
23340067 
24340054 
24540271 
and 15K05079 
), the Grants-in-Aid for Scientific Research on Innovative Areas from MEXT 
(No. 2404: 24105001, 24105008),
and by the Yukawa International Program for Quark-Hadron Sciences.

\end{document}